\documentstyle[12pt, graphics]{article}
\title{Finite Temperature Resonant Tunneling \\
in False Vacuum Decay and\\
the Lee-Yang Theorem}
\author{Laura Mersini\\
\it{Department of Physics}\\
\it{University of Wisconsin-Milwaukee}\\
\it{Milwaukee, WI 53201}\\
lmersini@uwm.edu}
\begin{document} 
\date{}
\maketitle
\begin{abstract}
\footnotesize
We consider the cosmological model of a self-interacting $\phi^4 - \phi^2$
quantum scalar field and extend our previous results, [3], on resonant
tunneling and consequent particle production, to the case of finite
temperature.

Using the mathematical equivalence between, the Euclidean path integral of
a $\phi^4 - \phi^2$ quantum field theory (in the saddle point
approximation), on one hand, and the partition function of a 4-dimensional 
ferromagnet (in the Ising model approximation), on the other, we derive 
the following results.

Tunneling is a first order  phase transition. The creation of metastable bound states
of instanton-antinstanton pairs under the barrier ,(i.e. resonant
tunneling), is the seed that gives rise to particle production. Through
the application of the Lee-Yang theorem for phase transitions, (as well as
demonstrating the underlying connection this has with the poles of the S-matrix element
in the quantum scattering theory), we show that the fluctuations around the
dominant escape paths of instantons (i.e. fluctuations of the bubble wall) with momenta comparable to the scale curvature of the bubble, drive  the mechanism for resonant tunneling
in false vacuum decay. We also identify the temperature dependence of the
parameters in the potential term, (or equivalently, of the instanton
bubbles), for a wide range of temperatures, including the critical
exponent near the critical temperature.(This is the temperature where the
1st order phase transition ends and the 2nd order transition begins).
Finally, we show  that the picture of a dilute instanton gas, which is
commonly used to describe tunneling in vacuum decay, remains valid even at
finite temperatures, as this gas becomes more and more dilute with the
increase of the temperature. This suppression continues until we reach the
critical temperature, at which point there is only one instanton left,
with an infinitely thick wall. 
\end{abstract}
\renewcommand{\thesection}{\arabic{section}.}
\renewcommand{\theequation}{\thesection\arabic{equation}}
\pagebreak
\section{Introduction}
\setcounter{equation}{0}
\vspace*{0.3cm}
\noindent
In this work we extend our results for resonant tunneling
and particle creation [3] to the finite temperature case.  The impetus 
came from two sources.Firstly,the realization that the dilute gas of instantons
can be described as an ensemble of paths (dominant escape paths, DEP) in the
trajectory space parameterized by the coordinate of the bubble wall 
$s$ (the time-like coordinate). The
time-evolution of the system in this parameter was equivalent to summing up
over all the members of the ensemble i.e. instantons.  Secondly, the analogy
with the theory of ferromagnets in statistical physics.  The mathematical
equivalence of the equations, particularly the partition function of ferromagnets
with the Euclidean path integral of the tunneling field in the saddle point
approximation allowed us to translate the well known results obtained for
ferromagnets in literature, to the case of finite temperature, instantons
dilute gas and the fluctuations of the scalar field around its DEP's.  \\
\\
\noindent
In Sections 1.1 and 2 we give a brief overview of the results for false
vacuum decay via tunneling and the theory of ferromagnets respectively as
a background material of Section 3.  In Section 3 we draw the conclusions
for finite temperature tunneling and particle production by exploiting the
insight given by 2 and the analogy of Sections 1.1 and 2.   In Section 4
we summarize these conclusions.  \\
\\
{\underline{\bf 1.1. Tunneling and Particle Production in Vacuum Decay.}}\\ 
\noindent
In our previous work [3] we calculated particle production in false vacuum decay,
via tunneling.
There the potential considered was (Fig.1.1)\\
\\
\begin{equation}
V(\Phi) = \frac{\lambda}{4!} (\Phi^2 - a^2)^2 - \frac{\epsilon}{2a}(\Phi + a)
\;\mbox{ where}\;  \frac{\epsilon}{\lambda a^4} \ll 1.  \label{potential}
\end{equation}
\vspace{1.0in}\\
Fig 1.1:

We split the field $\Phi$ in two parts, a classical field $\sigma$ (that is
the one that extremizes the action) and a fluctuating field $\varphi$ around
the classical field.  We solve a functional Schrodinger equation and find
out the form of the functional $\psi[\sigma,\varphi]$.  That is the usual
Gaussian form for saddle point approximation, i.e. the functional which is
a 1-parameter family of solutions and is peaked around the classical field
$\sigma$ in both vacuums.
Under the barrier $\sigma$ interpolates between 2 vacuum solutions, 
i.e. $\sigma$ = - a  $tanh[\mu(\rho_E - R)]$ (the usual instanton solution, [4])
where the parameter $\rho_E$ is the
4-euclidean radius and $R$ is the radius of the bubble at the point of
nucleation.  $\mu = \frac{1}{L}$ is the inverse of the thickness of the
wall.  We solve for the fluctuation field $\varphi$ also, and find
that we have resonant tunneling and particle production.  To solve for
the fields under the barrier we use Neumann boundary conditions.
The particle production rate was calculated by analytically continuing the
$\varphi$-solution from under the barrier, through the turning point
(at $\rho_E = 0$) and comparing it with what should have been a  positive
frequency wave in the true vacuum region.  The parameter ``$\rho_E$'' 
played the role 
of a time-variable under the barrier and letting it run from 0 
to $\infty$ was equivalent to summing up over all instantons in the dilute
gas approximation $(\mu R \gg 1)$.  Thus the ``time-evolution'' of the
field is equivalent to the trajectory over the ensemble of paths (DEP's) in 
configuration space.  That is because instantons are DEP's, in
configuration space and it is helpful to think of it as a phase space.\\
\\
The fluctuation field $\varphi(\rho)$ around the DEP, i.e. around the field
 $\sigma(\rho)$, gives
rise to particle production. This is basically due to the fact that the
phase shift in the functional which arises from the term 
$V''(\sigma)\cdot \varphi^2$ changes from imaginary to real while going
from Euclidean (under the barrier)
to Minkowski region (true vacuum). We calculate the particle production number
 from the fluctuation field
$\varphi$, taking into account the dynamics of the background tunneling
field $\sigma$.In that paper [3], the particle production number $n_p$ was found to be 
\begin{eqnarray}
\;\;\;\;n_p &\! =\! N & \frac{ |c_2|^2}{|c_2|^2 + |c_1|^2}\\
& \! =\! & 
\frac{4 \omega^2 q ^2 \cos^2 (\Delta^2 {\sqrt{-\alpha_3}}) - 
\pi^2 e^{- 2\pi p - 2 \theta_3} \cdot \sin^2(\Delta^2 {\sqrt{-\alpha_3}})\cdot [\omega^4 + q^4_]}
{4(\omega^2+q^2)^2 \cos^2(\Delta^2 {\sqrt{-\alpha_3}}) + \sin^2(\Delta^2 {\sqrt{-\alpha_3}})\omega^2q^2}\nonumber.\label{pnumber}
\end{eqnarray}  

where; {${\omega}, q,{\Delta},{\alpha_3}$} are functions of the momentum p, the inverse bubble thickness $\mu$ and nucleation radius R. $N$ is a normalization prefactor. (see [3] for the explicit calculation and dependence form  of the above parameters. Also, the coefficients $c_1$ and $c_2$,in the quantum scattering context, correspond to the reflection and transimission coefficients found from the solution of the Schrodinger equation in the Euclidean region.)
 We found a new feature, namely, resonant tunneling and
particle production.  Thus particle production was enhanced in comparison
to calculations of [M.Sasaki, et al. [5] ].
The resonant peaks in the particle production spectrum were located at
values of momentum multiples of the scale-curvature of the bubble, i.e.
$p \simeq n \mu R$ and they obeyed the exponential suppression feature, i.e.
the height of peaks decreased with the increasing values of momentum $p$ (Fig. 1.2).
In Section III we explain the resonant tunneling and particle production by the analogy with ferromagnetic systems and the application of the Lee-Yang theorem.\\
\\
\section{Brief Review.  Theory of Ferromagnets in Statistical Mechanics}
\setcounter{equation}{0}
\underline{\bf 2.1. The Analogy between QFTh and Statistical Mechanics Systems}\\
\\
For a field theory governed by a Lagrangian $\cal L$, the
generating
functional of the correlation function [1], (Schwinger function e.g.) is
\begin{equation}
Z[\Gamma] = \int D \varphi \exp[i \int d^4 x ( {\cal L} + J \varphi)]. \label{partition}
\end{equation}
The time variable of integration in the exponent runs
from $-T$ to $T$ with $T \longrightarrow \infty (1 - i \epsilon)$.
Correlation functions are reproduced by:
\begin{equation}
< 0 | T(\varphi(x_1) \ldots \varphi(x_n)) | 0 > = Z^{-1}[J]
\left( \frac{- i \delta}{\delta J (x_1)}  \cdots \frac{- i \delta }{\delta J(x_n)}
\right)Z[J].
\end{equation}
This generating functional is reminiscent of the partition function of 
statistical mechanics.
It has the same structure of an integral over all possible configurations
of an exponential statistical weight.  The source $J(x)$ plays the role 
of an external field (in general any linear term in $\cal L$).  This analogy
is made more precise by manipulating the time variable of integration in
(\ref{partition}).  Like the Wick rotation in the momentum integral, this Wick rotation of the
time coordinate, $t \longrightarrow - ix^0$, produces an Euclidean 4-vector
product:
\[ x^2 = t^2 - |\stackrel{\rightarrow}{x}|^2 \longrightarrow - |x_{\overline{t}}|^2
= -x^{0^2} - |\stackrel{\rightarrow}{x}|^2 . \] 
It has been shown that the analytic continuation of the time variables in any
Green's function with $0(4)$ symmetry in quantum Field Theory (QFTh hereafter) produces a correlation
function invariant under the rotational symmetry of 4-dim.
Euclidean space.  We illustrate this Wick rotation for the $\Phi^4$-theory.
The action of the $\Phi^4$-theory coupled to a source $J$ is: 
\begin{equation}
\int d^4 x ({\cal L} + J \phi) = \int d^4 x \left[ \frac{1}{2} (\partial_\mu \phi)^2
- \frac{1}{2} m^2 \phi^2 - \frac{\lambda}{4!} \phi^4 + J \phi \right] .
\end{equation}
After the Wick rotation, (2.3) becomes:
\begin{equation}
\quad i \int d^4 x_E ( {\cal L}_E - J \phi)  =  i \int d^4 x_E \left[ \frac{1}{2}
\left( \partial_{E\mu}\phi \right)^2 + \frac{1}{2} m^2 \phi^2 + \frac{\lambda}{4!}
\phi^4 - J \phi \right] .
\end{equation}
The expression (2.4) is identical in form to the expression for the Gibbs free
energy of a ferromagnet in the Landau Theory, namely:
\begin{eqnarray}
H & = & \int d^3 x \left[ \frac{1}{2} (\bigtriangledown s)^2 + b(T)s^2 + cs^4 - hs \right]
\end{eqnarray} 
where $s(x)$ is the fluctuating spin field.  Hence the quantum field
$\phi(x_E)$ identifies with $s(x)$, the external magnetic field $h$ with
$J$ (or in (1.1) with $\frac{\epsilon}{2a}$), $c$ with $\frac{\lambda}{4!}$, and $b(T)$=$m^2$  with
$- \mu^2 = - \lambda a^2$ in (I.1).  Note that the QFTh ``ferromagnet''
lives in four rather than three dimensions.
The Wick rotated generating functional $Z[J]$ becomes:
\begin{eqnarray}
Z[J] & = & \int D \phi \exp [- \int d^4 x_E ( {\cal L}_E - J \phi) ].
\end{eqnarray} 
In this new form $Z[J]$ is precisely the partition function containing the 
statistical mechanics of a macroscopic system, described by treating the 
fluctuation variable
as a continuum field.  We do the same, in the instanton language of the
tunneling problem by summing over the ensemble of the dilute instanton gas.
>From the above prescription,
 the Schwinger function in QFTh becomes correlation function
in statistical mechanics.  Thus we can say that fluctuations in QFTh behave
statistically.  \\
\\
It is well known that the qualitative behavior of a QFTh is not determined by
the fundamental Lagrangian but rather by the nature of the renormalization 
group flow and its fixed points[1].  These, in turn, depend only on the basic
symmetries that are imposed on the family of ``$\cal L$'' that flow into
one-another. Exploiting the analogy between statistical mechanics and QFTh,
it has
been shown that there is a strong relation between critical exponents of
statistical mechanics and the running of coupling constants to their fixed 
points in
QFTh.  This relation of critical exponents with the renormalization group equations (RGE) will shed light into one
of the questions we pose later, namely:  What's the temperature 
dependence of correlations and coupling constants during phase-transition
from one vacuum to another and their behaviour while approaching the critical
temperature?  \\
\\
Firstly, we will briefly summarize the Landau's Theory of phase transitions.\\
\\
{\bf II.2 Landau Theory of Phase Transition.}\\
\\
In thermodynamics a 1st-order phase transition is a point or a boundary 
across
which some thermodynamic variable changes discontinuously.  At a phase
transition point, two distinct phases are in equilibrium and the variable
that changes discontinuously is called the order-parameter.  If for some
value of the parameters the two phases become identical (i.e.there is one unique phase),
the discontinuity disappears. That is the critical point which 
signals the start of a 2nd-order transition [1,2].  \\
\\
At this critical point, one thermodynamic phase bifurcates into two distinct phases of the system. 
 It
is the coalescing of the two phases that leads to the long-range thermal
fluctuations which,beyond this point dominate over all other fluctuations.  An example
of this is the behaviour of a ferromagnet described by the Hamiltonian (2.5).
The total magnetization 
\begin{eqnarray}
M & = & \int d^3 x < s (x) > 
\end{eqnarray}
is the order-parameter, and $s(x)$ 
is the fluctuating spin wave. 
At low temperature, application of an external field $h$ will favour one of
the two possible phases.  At $h = 0$, the two phases are in equilibrium. 
 If $h$
changes by a small amount from negative to positive values, $M$ changes
discontinuously (Figure 2.1).  \\
\vspace*{2.0in}\\
\centerline{\bf Figure 2.1}
Thus for any low temperature there is a phase transition at $h = 0$.  
When the temperature is raised, the fluctuation of spin increases and the
value $|M|$ decreases.  At some critical temperature $T_C$ the system
ceases to be magnetized, at $h = 0$.  At this point, the two phases coalesce
and that is the end of the 1st-order transition, as in Figure 2.2.\\
\vspace{2.0in}\\
\centerline{\bf Figure 2.2}
Landau described this behaviour, by the use of the Gibbs free energy $H$,
which depends upon $M,T$ such that:
\begin{eqnarray}
\left. \frac{\partial H}{\partial M} \right|_T   & = & -h 
\end{eqnarray}
He suggested that if we concentrate near a critical point, 
$T \approx T_c, M \approx 0$, we can expand $H(M)$ in a Taylor series.  For
$h \cong 0$ we can write
\begin{eqnarray}
H(M) & = & A(T) + b(T) M^2 + cM^4 + \ldots
\end{eqnarray} 
Because of the symmetry $M \rightarrow -M$, for $h = 0, H(M)$ must have even
powers of $M$ only.  By minimizing the energy he found:
\begin{eqnarray}
b(T) & = & b_0 (T - T_c), \quad c(T) = c \nonumber \\
& & \\
M & = & \left\{ \begin{array}{cl}
0 & T > T_c \nonumber\\
\pm \left[ (\frac{b_0}{2c}) (T_c - T)^{1/2}\right] & T < T_c\; .
\end{array} \right. \nonumber
\end{eqnarray}
Then, the Green's function $D(x)$ is:
\[ D(x) \sim <s(x) s(0) > = \sum_{\mbox{all}\, s(x)}s(x)s(0)e^{- E/kT} \]
where $E$ is the microscopic Hamiltonian of the magnetic system.  By ignoring the
$s^3(x)$ in the equation of motion, near $T_c$, he calculated [1,2]:
\begin{eqnarray}
D(x) & = & \frac{h}{4\pi r} e^{- r/\xi} \nonumber\\
& & \\
\xi & = & [2b_0 (T_c - T)]^{- 1/2}, \; \mbox{correlation length.} \nonumber
\end{eqnarray}  
The correlation length, $\xi$, diverges
as $T \rightarrow T_c$.  The power-law dependence of correlations on
$(T_c - T)$ is universal for many systems.  In the case of a $\Phi^4$-QFTh,
the coefficient in the power law is shown to be $1/2$, exactly the same
as in the 4 dimensional ferromagnet systems.\\
\\
{\underline{\bf 2.3 Correlation Inequalities, Lee-Yang Circle Theorem and}}\\
{\underline{\bf other Related Theorems.}}\\
\\
The simplest inequalities are the Griffith's inequalities[2] which states
that expectation values and pair correlations are positive for general
ferromagnetic interactions without the quartic term.  The Griffith's inequalities have been used
in the Ising model study of ferromagnets and it has been shown that the
correlation functions converge in the limit of infinite volume.
>From the equivalence of the statistical mechanics models considered in (2.5) and QFTH models (2.6) 
 we can
use the results in the study of finite-temperature phase transition in the
QFTh.  Note that the Ising model is a lattice based approximation but the
continuum limit of QFTh is achieved by taking the volume to infinity.\\
\\
The proof of the Lee-Yang Theorem is based on Griffith's inequalities [2].The Lee-Yang theorem (referred below as $Theorem I$) states that there is a phase transition occurring when the partition 
function becomes zero at $h = 0$.\\
\\
An enhancement of the Griffith's inequalities are the FKG inequalities which
allow for the inclusion of interactions, specifically quartic terms.  They
include interactions by using the mean-field theory and allowing for
fluctuations around the mean-field.  That is exactly the semiclassical
saddle point approximation in QFTh, where the mean-field corresponds to the 
classical field.  The condition for FKG inequalities to hold is that all
 higher
derivates of the potential should be much less than the 2nd-derivative.
This condition translates in QFTh to the WKB condition that the potential
barrier that separates the two vacua (phases) should be sufficiently high
and wide.  The generalized Lee-Yang Circle Theorem, based on FKG inequalities
states that:  a phase transition occurs whenever the partition function
 $Z[h]$
becomes zero at some real value of the parameter $h = h_0$.  In this
case, the phase transition may not occur for the 0-th order values of the potential calculated in the mean-field approximation, i.e. at $V(\Phi) = V(\sigma_{cl})$
but it may happen for the  perturbed values of the potential which allow the inclusion of the fluctuations $\varphi$ around the mean-field (classical field) $\sigma_{cl}$, i.e. at  values:\\
\\
\begin{equation}
V(\Phi) = V(\sigma_{cl}) + \delta V, \; \mbox{with}\; \delta V = 
\frac{1}{2!}V''(\sigma_{cl})\cdot \varphi^2    
\end{equation}
Thus, the Lee-Yang Theorem can be readily used in the $\Phi^4$-tunneling
case in QFTh.  The Lee-Yang Theorem shows that for systems described by (2.5)
there is a phase transition for $b(T) < 0$ at $h = 0$.  From
Figure (2.1), there are two distinct phases and a jump discontinuity of
$M(h)$ at $h = 0$.  Hence, if $b(T) > 0, \; M(h)$ is a smooth function
 of $h$
and there is no phase transition (Figure 2.3).\\
\vspace*{2.0in}\\
\centerline{\bf Figure 2.3}
The critical spacetime dimension for phase transition to occur is $d_{cr} = 2$,
i.e. it can happen only if $d > d_{cr}$.  This circle of ideas applied
to $\Phi^4 - \Phi^2$ QFTH shows that the classical (zero temperature)
ground state is defined by a constant field configurations.  The dominant
contributions to the quantum ground state (positive temperature) are close
to the classical one because of the large factor: $\exp(-S_E)$ (statistical
weight) but they contain various fluctuations.  The large scale fluctuations
are the tunneling transition between distinct wells and they are described
by the Ising fluctuations for $n=1, \; d=4$.  $n$ is the spin dimension
(i.e. a scalar spin wave), $d$ is the spacetime dimension.  It also means
that phase transition can occur only if $d > 2$.(Lee-Yang theorem is proven to hold for dimensions up to 4. It gives the same result as the Landau theory of the previous section when the ferromagnet is 4 dimensional)\\
\\
Thus, we are able to draw results for finite temperature tunneling and the  particle production
associated with it, in QFTH. It is important to note that the free Helmoltz energy per unit of 
volume $\Lambda$ is 
\[ A_{\Lambda}(h) = \frac{\ell n Z [h]}{\Lambda}. \]
That is finite when $\Lambda \rightarrow \infty$ with the condition of
using Neumann Boundary conditions.  Also
\begin{eqnarray}
\frac{dA_\Lambda}{dh} & = & < \Phi (x_E) >_h = M(h) = \sigma_{cl}\\
\chi = \frac{d^2 A_\Lambda}{dh^2} & = & \frac{1}{k_BT} \int (< \phi(x_E)\phi(y_E)>
- < \phi(x_E) > < \phi(y_E) > dy = \nonumber \\
& = & \frac{1}{k_B T} \int D(x,y)dy \nonumber
\end{eqnarray}
where; $h$ is the external field, $\chi$ is the susceptibility and $D(x,y)$ is the 2-point function.  
Thus the
first relation in (2.13) shows that the classical field $\sigma$, is the
magnetization, while the second is simply the ``fluctuation-dissipation''
relation.  Obviously the equivalent of $A(h)$ in QFTH is the effective action.
It is important to remark that symmetry breaking and phase transition are
two distinct phenomena, although, in general often people use them interchangeably.
We will show this distinction when we connect and apply the above ideas to
resonant tunneling in vacuum decay. Before that we will introduce another theorem,
based on that of the Lee-Yang one [2], which will be needed later.\\
\\
{\bf Theorem (II):} {\it Consider a $\phi^4$
or Ising model with zero 
external
field.  The Hamiltonian (2.5) restricted to ${H}_{\mbox{even}}$
(the subspace of $ H$ invariant under $\Phi \rightarrow - \Phi$
isomorphism) has no particle spectrum for $b > b_c,$ or equivalently
there are no even bound states with energy below the 2 particle threshold.}\\
\\
Near the critical point $b_c$, such as $b > b_c$, all the derivatives of the
correlation function $D_n$ exist.  Their behaviour is given by the critical
exponents (fixed points of RGE) which for the Ising model under 
consideration $(n=1, d=4)$ and large separation $|x|$, are as follows:
\begin{equation}
<\phi(x)\phi(0)>| \sim \frac{h}{4 \pi |x|} e^{-m|x|} \quad
m = \xi^{-1} \sim (b_c - b)^{1/2} = [b_0  (T_c - T)]^{1/2}
\end{equation} 
This result was also obtained by Landau (Section 2.2) 
\section{Finite Temperature Tunneling and Resonant Particle Production, in
Vacuum Decay in QFTh}
\setcounter{equation}{0}
From the description above (Section 1.1 and 2), we now have in place the information we need to gain an
insight into finite temperature tunneling, using the
equivalence of the Ising model in statistical mechanics and tunneling 
in vacuum decay in QFTH.\\

{\bf 1)} Tunneling is a first order   phase transition.  While there is an explicit symmetry breaking for the classical field $\sigma$ for $\epsilon \neq 0$ in Eqn. (1.1), there is no phase transition
at that point, if we restrict the treatment to the classical field and  don't include the fluctuations, as we now know from the Lee-Yang theorem.  Note that
$\epsilon$ in equation (1.1)acts as an ``external magnetic field''.
From the Lee-Yang theorem restricted to the classical field only, transition occurs only when 
$\epsilon \rightarrow 0$.
It is only at that point that $Z[\epsilon]$ becomes zero.  But for
$\epsilon = 0$ the 2 phases (vacuum states) are identical, so there is no
symmetry breaking in the dynamic sense of the term.However, if we take the fluctuations into account there is a phase transition occuring,even when the 'external magnetic field' is zero, i.e. when  $\epsilon$=0.Thus, in the absence of an 'external field', the phase transition is rendered  by the fluctuation field \\
\\
For $\epsilon \rightarrow 0$, the Hamiltonian in (1.1) is even under the
exchange $\phi \rightarrow - \phi$, thus there must be particle creation
(and indeed there is) according to Theorem (II).This resonant particle production is due to the
creation of even metastable bound states.\\

{\bf 2)} These bound states are pairs of instanton-antiinstantons. They
are metastable bound states and give rise to a particle spectrum, as per
(II).  The statistical mechanics picture of the instanton dilute gas 
and instanton-antiinstanton pairs is the dipole gas condensation.  
Further insight is gained by thinking
of, yet another picture, scattering theory and partial wave analysis.  
In that
picture, it is well known that the creation of even metastable bound
states indicates resonance, and those, (and the corresponding phase shifts
$\delta_\ell$) are found by the poles in the $S$-matrix (the 2-point function
 in our
approximation) for imaginary values of the momentum.  Thus the two pictures
are revealing that they are,in fact, equivalent.  Namely the application of
the Lee-Yang Theorem and finding the poles of the $S$-matrix.  A pole of 
the $S$-matrix
in the imaginary momentum (the Euclidean region) is equivalent to the partition
function becoming zero, since the latter goes as the exponential of
the negative 2-point function (in our WKB approximation).  So, for the phase 
transition to
occur we need poles of the $S$-matrix.  This implies  that first order phase transition
occurs through resonant tunneling (and consequently resonant particle
production)[3].  The metastable bound states of the instanton- antinstanton pairs 
( corresponding to vortices or dipole gas condesates  in a statistical mechanical system)are the mechanism that makes it possible. This connection is
very useful since we can find the phase structure of tunneling in real time
by studying the poles and phase shifts of the $S$-matrix through scattering
theory. As a simple application of this method we can make use of the result (1.2) found in a previous paper [3]. There, in order to calculate the particle production number $n_p$, we needed to solve a Schrodinger equation under the barrier, i.e. in the Euclidean region. $n_p$ was expressed as a function of  the two coefficients, $c_1$, $c_2$ that were found through solving the Schrodinger equation and matching the solutions. Those two coefficients were the reflection and transmission coefficients respectively, in the language of quantum scattering theory ( see Eq. (3.17-3.19) of paper[3] for the details). Thus a pole in the $S$- matrix element corresponds to the reflection coefficient $c_1$ becoming zero. But this coefficient $c_1$ becomes zero at values of the momentum p which are also  the resonant peaks of $n_p$ (Fig.1.2) i.e. at $p \sim  n \mu R$. 

 This also reinforces the deep relationship between tunneling and
particle production.  The identifications made by comparing (1.1) with (2.5) (most of which have already been mentioned
in Section 2) are given in Eq. (3.1-3.3) below
\begin{eqnarray}
\epsilon \longrightarrow h,  &&   \mbox{the external field}\nonumber \\
\sigma \longrightarrow M(h), &&  \mbox{instanton field is the magnetization} \\
&&                                \mbox{(order-parameter)}\nonumber \\
S_E \rightarrow A_\Lambda(h), \; \mbox{for} \; \Lambda \rightarrow \infty,&&
\mbox{free energy and the Euclidean effective action}.\nonumber
\end{eqnarray}          

{\bf 3)} Temperature Dependence of Tunneling.
From the equivalence of the action (1.1) to (2.5) we see that $-b(T)$
identifies with $\mu^2 = \frac{1}{L^2} = \lambda a^2$, where $L$ is
the thickness of the wall.  Thus from the Ising model and Landau critical
exponents (2.11) we can translate immediately the temperature
dependence of tunneling parameters as follows
\begin{eqnarray}
\mu^2 & = & \mu^2_0 ( 1 - \frac{T}{T_c}) \nonumber \\
a & = & a_0 \sqrt{(1 - \frac{T}{T_c})}\\
\sigma & = &\pm \sigma_0  \sqrt{( 1 - \frac{T}{T_c})} \nonumber
\end{eqnarray}
$\lambda$,like the coefficient $c$ in (2.5) with which it is identified does not depend on the temperature.Here,  $\mu^2_0 = \lambda a^2_0$.
Also from (2.14)
\begin{eqnarray}
D(x) \cong \frac{\epsilon}{4\pi|x|} e^{-\sqrt{2\lambda a^2_0({1-\frac{T}{T_c})}|x|}}
\cong \frac{\epsilon}{4\pi|x|}e^{ - \mu_0|x|\cdot \sqrt{1 - \frac{T}{T_c}} }
\end{eqnarray}  
Thus, from (3.2), the wall of the bubbles thicken as the temperature is
raised, until we reach the critical temperature $T_c$, at which point, the
location of vacuum is $a = 0$ (2 phases coalesce) and there is only one
disordered phase, the wall itself.  The critical temperature $T_c$ for
the Ising model is found to be $T_c = \frac{\mu^2_0}{2\pi}$.  This 
result coincides with the Caldeira and Leggett [6] calculation of the crossover
temperature $T_0$.  They use a different method, but in the saddle point 
approximation they find that the temperature where the transition from
quantum to classical occurs is: $T_0 = V_0/2\pi$ where $V_0$ is the
height of the barrier.  It is to be expected that the critical temperature
$T_c$, is the same as $T_0$ since the thermal fluctuations at $T_c$
become long range and dominate vacuum fluctuations.The temperature dependence of the particle production number $n_p(T)$ is readily obtained by replacing the temperature dependent parameters  of Eqn's (3.1-3.2) in the formula (1.2). (The explicit dependence of the coefficients $\omega$, $\Delta$,$q$,$\alpha_3$ on the above parameters is given in paper [3]).It is straightforward to check that $n_p(T)$ goes to zero when the temperature approaches the critical one,$T_c$, by replacing the parameters  $\mu$, $a$, and $\sigma$ with  zero.  \\

{\bf 4)} An important issue is the fact that the number of instantons is also
reduced as the temperature goes up.  Finite temperature effects are
incorporated by averaging the quantum tunneling rate $\Gamma(E)$ with
the canonical equilibrium probability.  So the quantum decay rate
becomes
\begin{eqnarray}
\Gamma(\beta) = Z^{-1}_0 \int^2_1 dE   \Gamma(E) \exp [- \beta E]
\end{eqnarray}
where $E$ is the energy, and $\Gamma(E)$,(the decay rate), is the usual
exponential of the Euclidean bounce action.  As it's clear from (3.4)
the instanton gas is suppressed by the factor $\exp[-\beta E]$.  Thus,
despite that the wall thickens as $T \rightarrow T_c$ (3.2), and becomes
infinite at $T_c$, the numbers of bubbles/instantons is decreasing
due to the exponential suppression in (3.4). Thus the instanton
gas becomes more and more dilute, until at $T = T_c$ there is only 1
bubble with infinite thickness.  The interior and exterior of the
bubble (false and true vacuum) have coalesced at $T_c$ to form one
disordered mixed phase (i.e. 
$a \rightarrow 0$ at $T \rightarrow T_c$).  At this point the
thermal fluctuations are long-range.  They dominate over other fluctuations
and there is no tunneling or resonance.  This is the point of 2nd-order
transition.  Thus the dilute instanton gas becomes more and more dilute
as the temperature goes up.  As in the Peirles model [2], of liquid-gas transition,
bubbles below critical size shrink gradually to zero, while above the
critical size they grow by condensing with the other
bubbles until they reach complete condensation where there is one
giant bubble that covers the whole space.  The same process  happens with instanton 'bubbles'.
\section{Conclusions}
\setcounter{equation}{0}
In the present work, we have examined the scenario of false           
vacuum decay through resonant tunneling and subsequent particle creation
at finite temperature, from the viewpoint of the analogy of Euclidean
$QFTH$ with the Ising model of ferromagnets in condensed matter physics.\\
\\
First, we gave a brief overview and summarized the results on WKB multidimensional
tunneling and resonant particle production found in previous work [3] as
well as quoting the main theorems and results,which are well established in
the literature [1,2] for the theory of ferromagnets.  \\
\\
It is important to note that once we increase the temperature, the particle
creation number is exponentially suppressed by $e^{-\beta S_E}$, despite
that the prefactor increases with temperature; thus, it is consistent to neglect in the  equations of motion for the tunneling field $\sigma$,  the coupling of the $\sigma$ field to the fluctuatuion field $\varphi$, even at finite temperatures up to  $T_C$. The reason, as given by the above argument, is that the fluctuation field 
$< \phi^2>_T$ remains small even at finite temperature due to the exponential suppression $e^{- \beta S_E}$.\\
\\
The insight we gained by the analogy between 4-dimensional ferromagnets
and tunneling in QFTH (or rather the mathematical equivalence of the
respective equations) lead us to the following conclusions.
\begin{itemize}
\item[{i)}] Tunneling is a first order  phase transition.
\item[{ii)}] Thus, it is always accompanied by particle creation, as the first order transitions are always associated with  a mass gap (defined as the inverse of the correlation length $\xi$ as in Eqn (2.14)), while in the second order transitions the mass gap vanishes(see theorem II). Since the mass gap decreases with the increase of temperature while approaching the critical point (from below), the particle production number due to tunneling (i.e. the one that arises as a result of quantum fluctuations around the DEP's) should also decrease. Quantitatively this result is clear from formula 1.2 for $n_p(T)$ with the temperature dependence of the parameters $\mu$, $a$,$\lambda$ and $\sigma$ given in (4.1) below.
\item[{iii)}] Lee-Yang Theorem for phase transitions is equivalent to finding the poles of the scattering matrix element
 in the sense that the corresponding partition 
function always becomes
zero when the 2-point function of QFTh (S-matrix element), calculated in 
the imaginary
values of momentum, becomes divergent.  This provides insight into the mechanism of
tunneling, particle creation and phase transition in an $S$-matrix language.
The instanton-antiinstanton pairs of the tunneling field $\sigma$ form
metastable bound states which give rise to resonant particle production,and 
thus the phase transition.  The usefulness of this realization stands in
the fact that we can study the phase structure of finite temperature
tunneling in real time by finding the poles in the $S$-matrix.(Of course, having the poles and subsequently the phase shifts, one could calculate the scattering cross-section from the optical theorem.As an aside, often the method of poles in the S-matrix is useful when one has little knowledge about the potential term in the lagrangian).  
\item[{iv)}] We obtained the temperature dependence of the coefficients in the
Lagrangian (1.1) by the analogy with (2.5).  They are as follows (see Eqn (3.1-3.2)and Fig. 3 for the potential as a function of temperature)
\begin{equation}
\left\{ \begin{array}{lll}
\mu^2 = \mu^2_0 (1 - \frac{T}{T_c} ) & \mbox{(square of inverse}\nonumber \\
& \mbox{thickness of the wall)}\nonumber \\
\\
a = a_0 \sqrt{( 1 - \frac{T}{T_c} ) }   & \mbox{(vacuum location)}\nonumber \\
& \\
\sigma = \pm \sigma_0 \sqrt{(1 - \frac{T}{T_c})} & \mbox{(tunneling field)} \nonumber \\
\\
\lambda \; \mbox{independent of T.} & \nonumber \\
\end{array} \right.
\end{equation} 
where $T_c \simeq T_0 = \frac{\mu^2_0}{2\pi}$ (critical temperature and crossover
temperature).  In principle one could calculate (4.1) by using the
RGE.  That would be quite an elaborate and long calculation.  Thanks to
the analogy between the two models described above, (respectively the semiclassical QFTh model of  vacuum decay and the ferromagnet), we can instead use these results 
from the Landau theory of critical exponents in this case.
\item[{v)}] Finally, despite that the wall thickness of each bubble increases
with temperature, their number is suppressed.  Thus, the dilute gas
approximation remains valid below and up to  $T_c$, until at the critical temperature there is only one
bubble left with an infinite wall thickness.  This is the point where 1st
order phase transition ends and the two phases have meshed together in a
disoriented phase.  During this process, bubbles with radius larger
than critical, continued growing and blending together into yet larger
bubbles thus their number is decreasing until there is one very large
(infinitely thick wall) bubble left.
\end{itemize} 
\vspace*{2.0in}
\centerline{\bf Figure 3.0}
\vspace*{1.0in}
\noindent
Acknowledgment:

I would like to thank Professor L.Parker, my advisor, for the invaluable consulting, advice and support he has constantly given me through our discussions in the weekly meetings. I would also like to thank Dr.A.Raval and K.Lockitch for their help .
\pagebreak\\
\centerline{\large{\bf References}}
\begin{itemize}
\item[{1.}] Michael E.Peskin, Daniel V.Shroeder,eds(1996), ``An Introduction to Quantum Field Theory'' (Addison-Wesley Publishing Company),ISBN 0-201-50397-2 
\item[{2.}] James Glimm,Arthur Jaffe,eds(1981), ``Quantum Physics.A functional Integral Point of View'' (Springer-Verlag New York Inc)
\item[{3.}] L. Mersini, Phys.Rev.D 59, 123521, (1999). (hep-th/9902127)
\item[{4.}] S.Coleman,Phys. Rev. D.16,1248 (1977); C.G.Callan and S.Coleman ibid.16 1762 (1977) \\
\item[{5.}] M. Sasaki,T.Tanaka, Phys.Rev.D.49,1039 (1994). Hamazaki,M.Sasaki,T.Tanaka,K.Yamamoto, Phys.Rev.D.53,2045 (1995)
\item[{6.}] A.O.Caldeira and A.J.Leggett, Ann.Phys. (N.Y.)149, 374 (1983)
\item[{7.}] L.Parker, Phys.Rev.D.183,1057 (1969)
\item[{8.}] L.Parker, Nature,261,20 (1976)
\item[{9.}] T.Banks,C.Bender, and T.T.Wu,Phys.Rev.D 8,3366 (1973)
\item[{10.}] J.L.Gervais and B.Sakita,Phys.Rev.D 16,3507, (1977)
\item[{11.}] T.Vashaspati, A.Vilenkin, Phys.Rev.D.37,898 (1988). T.Vashaspati, A.Vilenkin, Phys.Rev.D.43,3846 (1991)
\end{itemize}

\resizebox{6in}{6in}{\includegraphics{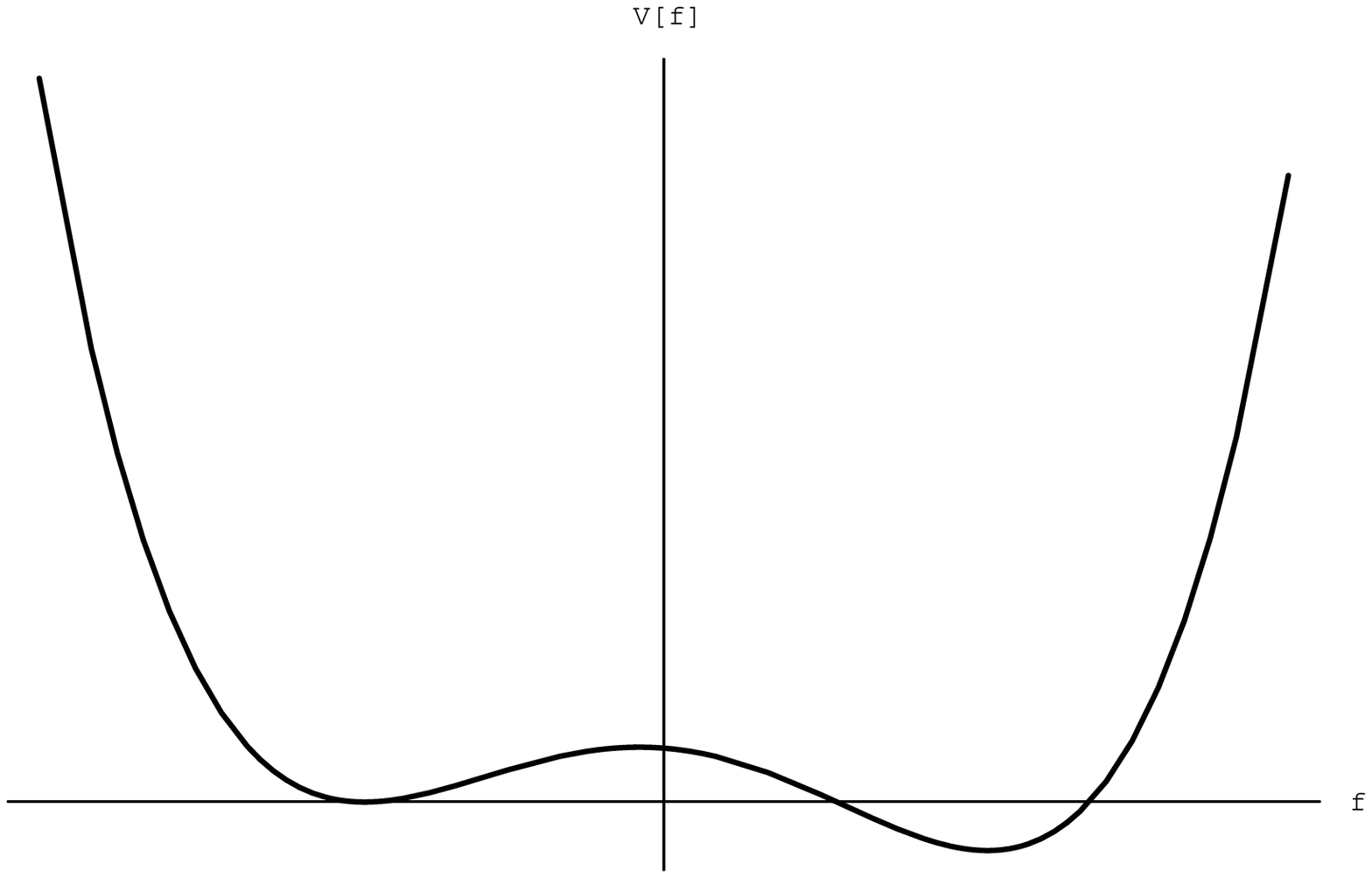}}
\begin{center}
Fig. 1.1 The potential for false vacuum decay via tunneling
\end{center}

\resizebox{6in}{6in}{\includegraphics{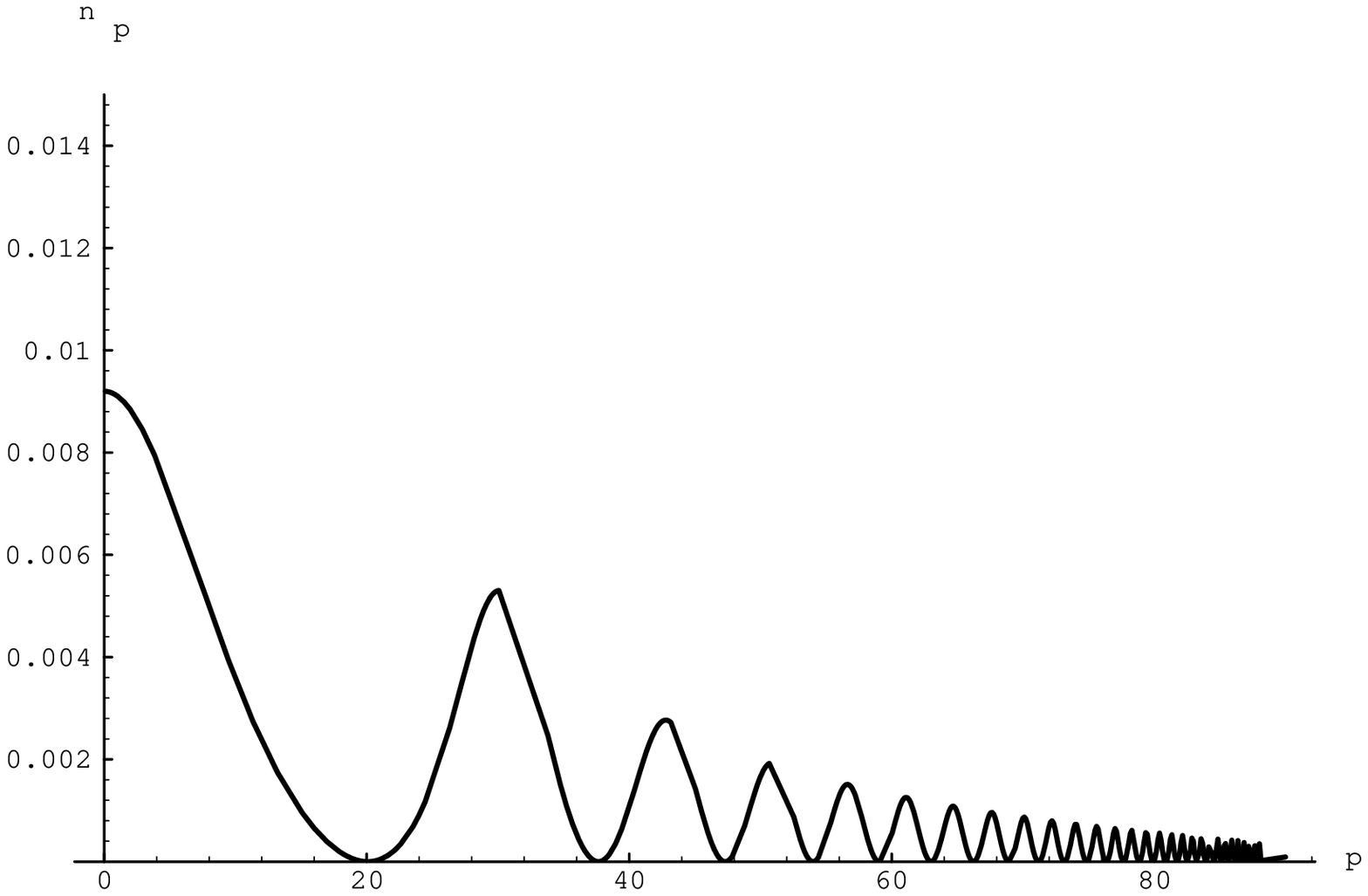}}
\begin{center}
Fig.1.2  Particle production number $n_p$ vs. p 
\end{center}

\resizebox{6in}{6in}{\includegraphics{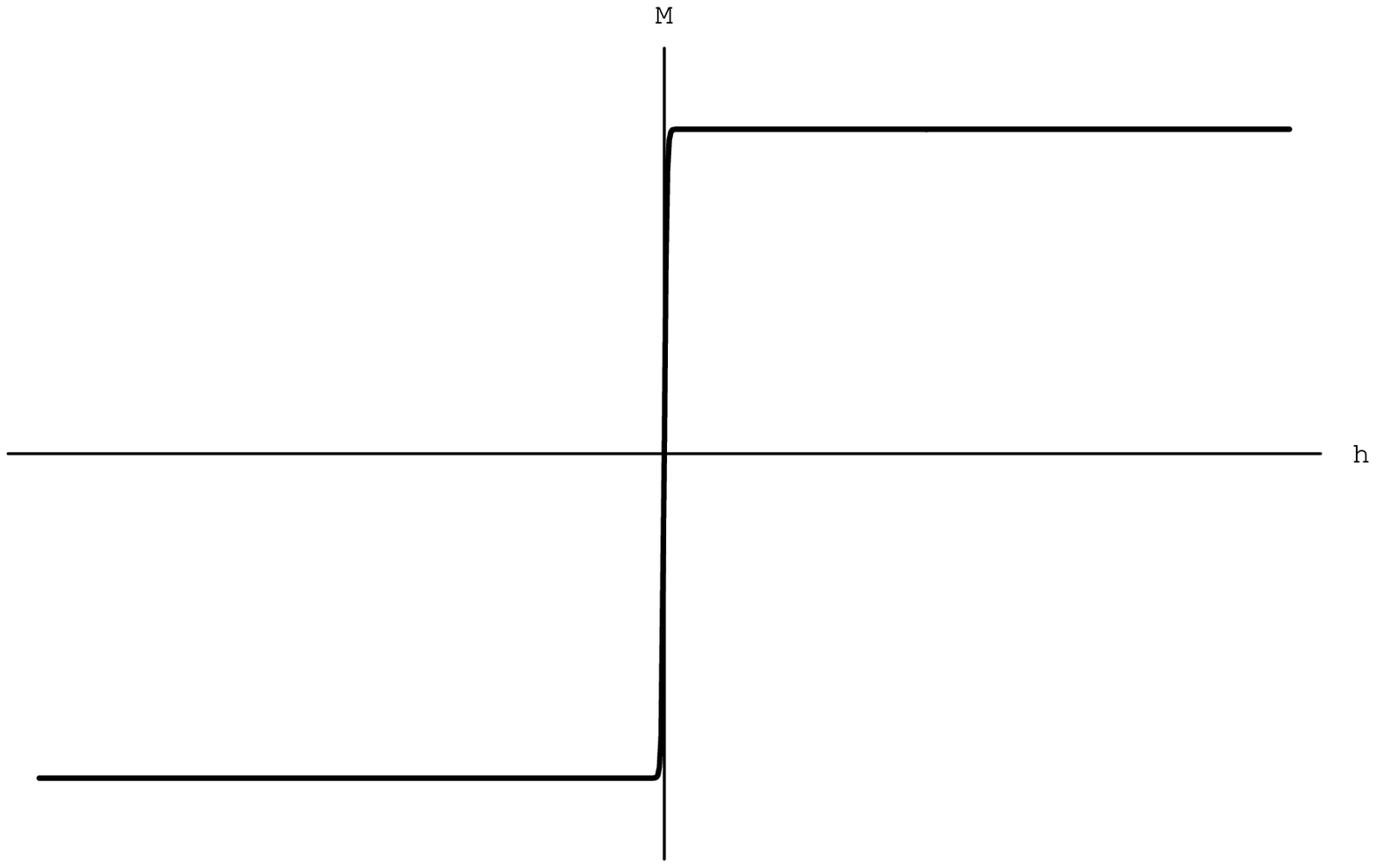}}
\begin{center}
Fig. 2.1  Magnetization M ($\sigma$) vs. the external field $h$ ($\epsilon$). The case of first-order transition
\end{center}

\resizebox{6in}{6in}{\includegraphics{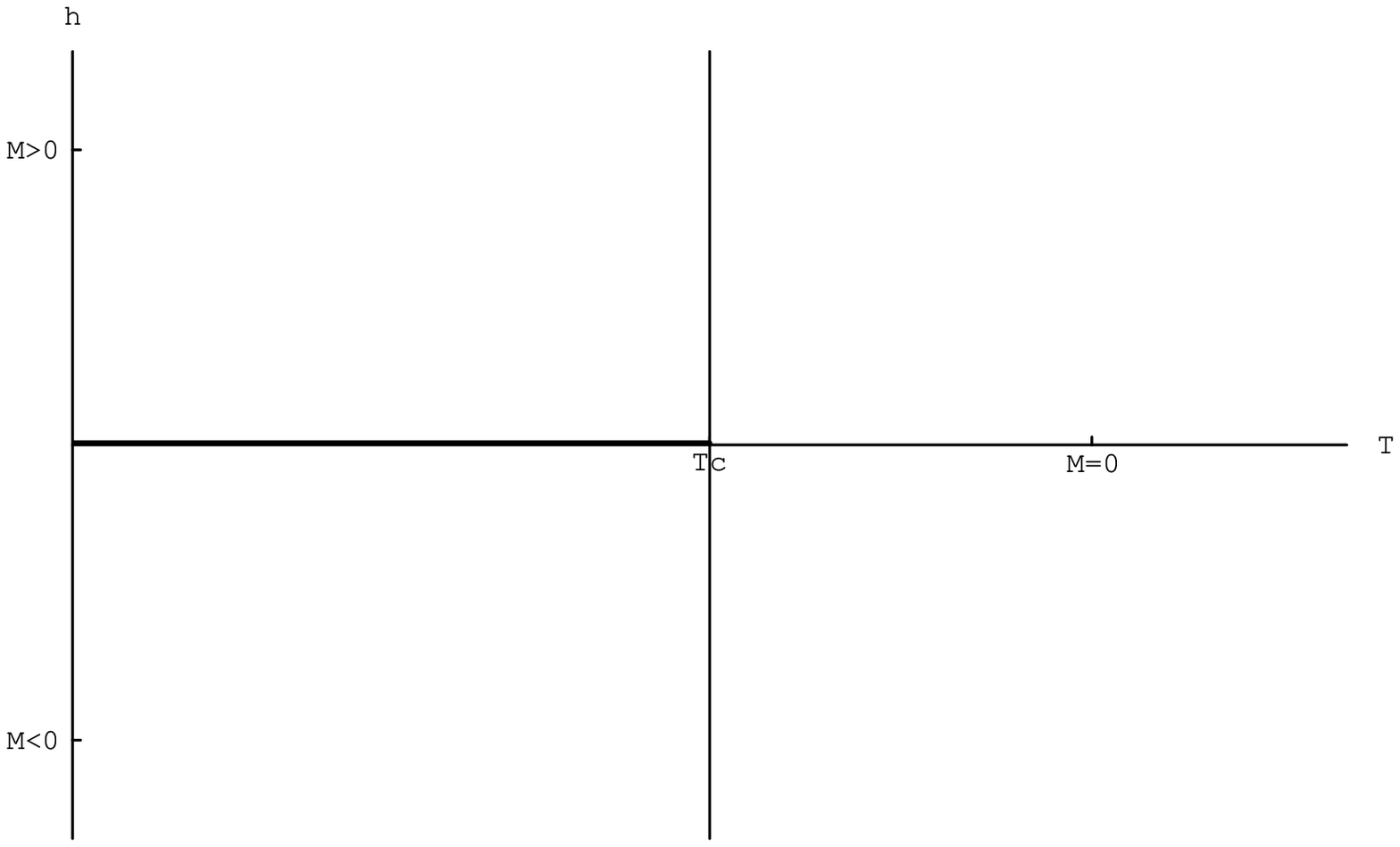}}
\begin{center}
Fig. 2.2  Phase transition as a function of temperature
\end{center}

\resizebox{6in}{6in}{\includegraphics{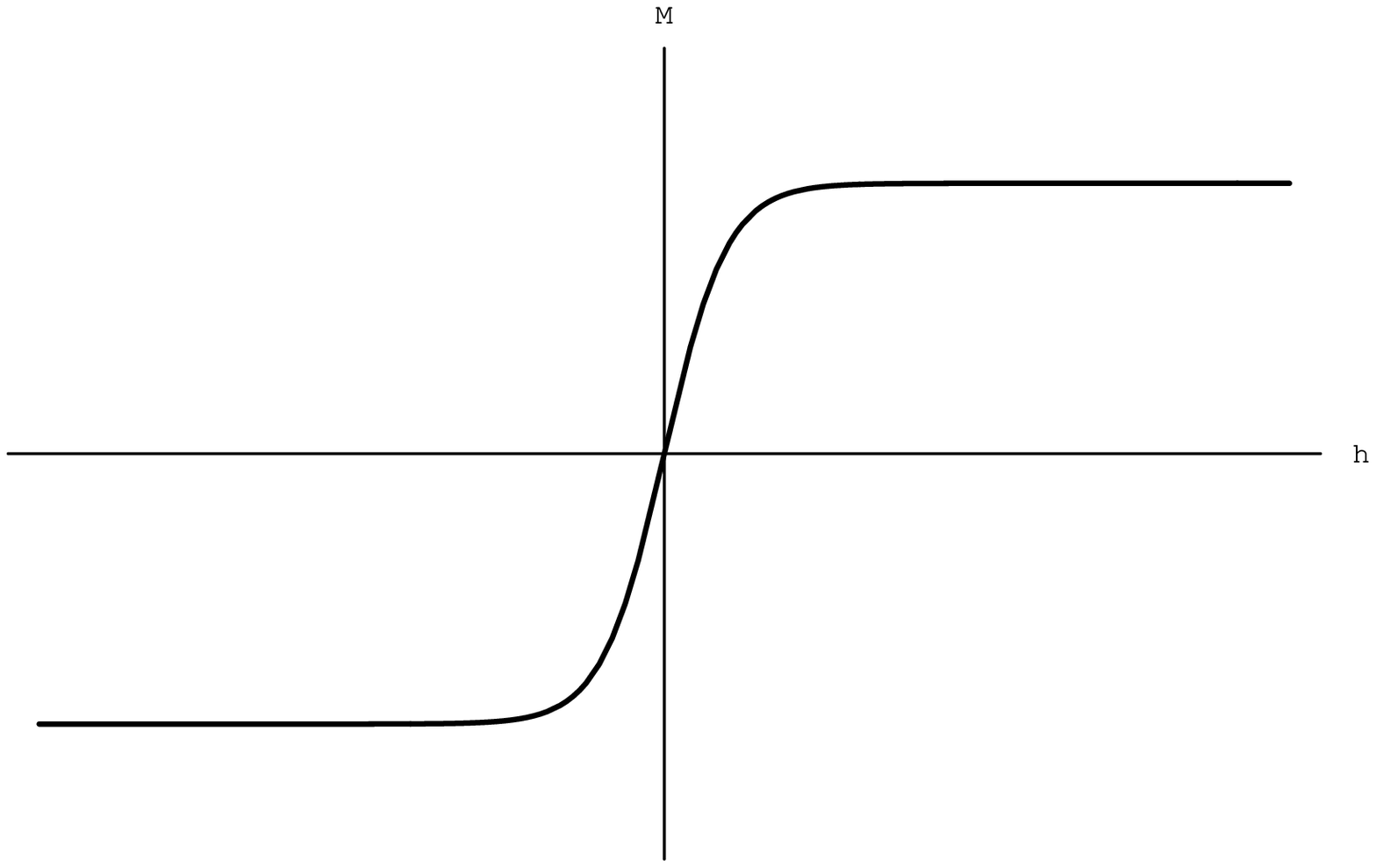}}
\begin{center}
Fig. 2.3  The case when there is no phase transition. Magnetization M is a smooth function of the external field $h$.
\end{center}

\resizebox{6in}{6in}{\includegraphics{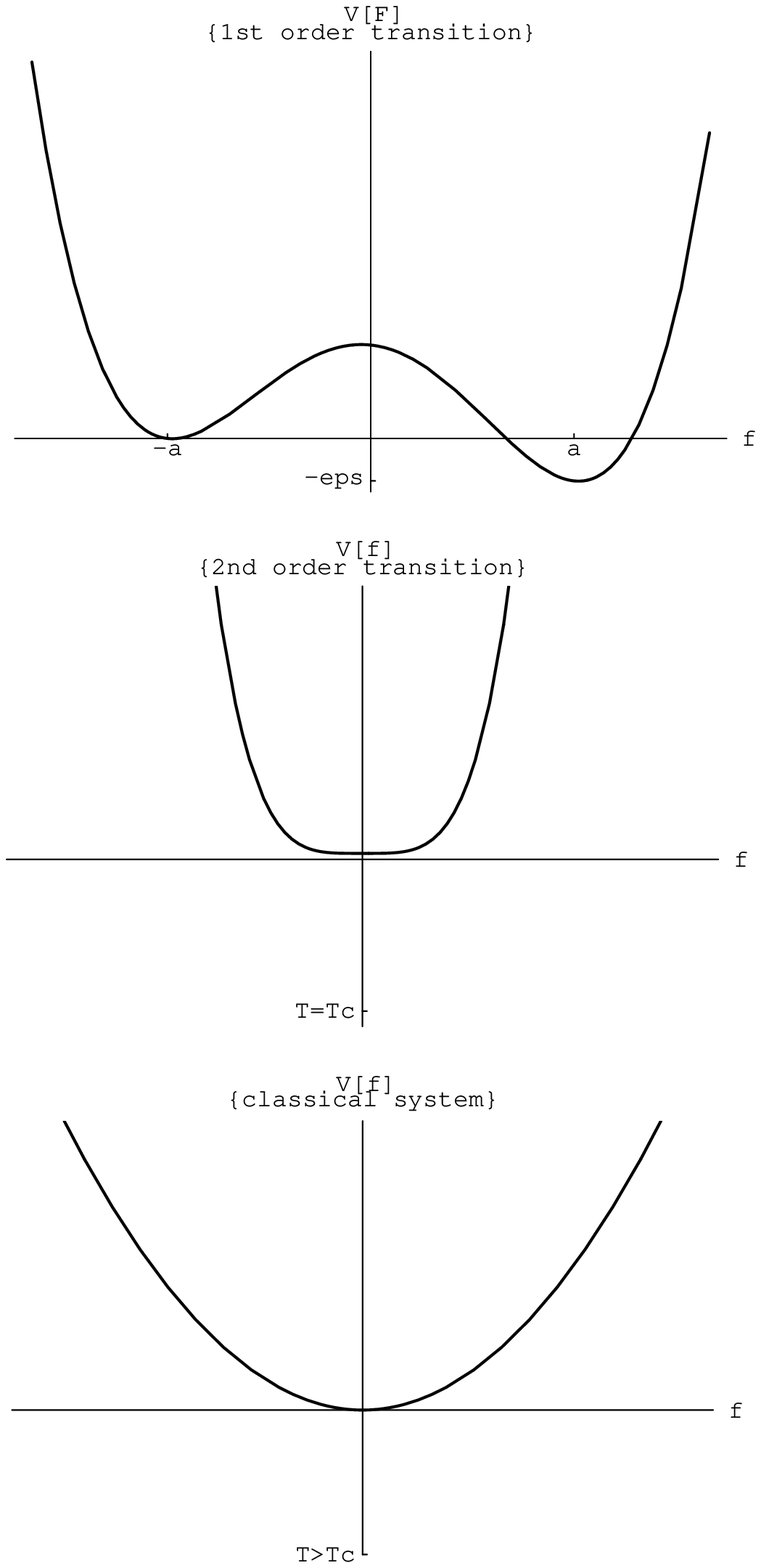}}
\begin{center}
Fig. 3.0  The potential of the scalar field at different temperatures (above and below $T_c$) 
\end{center}

\end{document}